\documentstyle[prl,aps,floats]{revtex}
\input{psfig.tex}
\tighten
\begin{document} 
\draft
\twocolumn[\hsize\textwidth\columnwidth\hsize\csname@twocolumnfalse\endcsname 
\title{Opportunities for future supernova studies of cosmic acceleration}
\author{Jochen Weller$^{1,2}$ and Andreas Albrecht$^1$ \\
$^1$ Department of Physics, University of California at Davis, CA
95616, U.S.A.\\
$^2$ Backett Laboratory, Imperial College, Prince Consort Road, London SW7 2BZ, U.K.\\
}
\maketitle
\begin{abstract}
We investigate the potential of a future supernova dataset, as might be
obtained
by the proposed SNAP satellite, to discriminate among 
different ``dark energy'' theories that describe an accelerating
Universe.   We find that
 
many such models can be distinguished with a fit to the effective
pressure-to-density ratio, $w$, of this energy.   More models can be
distinguished when the effective slope, $dw/dz$, of a changing $w$ is
also fit, but only if our
 
knowledge of the current mass density, $\Omega_m$, is improved. 
We investigate the use of ``fitting functions'' to interpret
luminosity distance data from supernova searches, and argue in favor
of a particular preferred method, which we use in our analysis.
\end{abstract}

\date{\today}

]

\renewcommand{\thefootnote}{\arabic{footnote}}
\setcounter{footnote}{0}
Increasing evidence that the Universe is accelerating \cite{BOPS}
confronts us with deep unresolved questions.
As yet no compelling
understanding of the acceleration has been achieved but many
models have been proposed, typically introducing a form of ``dark
energy'' to account for cosmic acceleration. 
Further resolving the observational evidence for acceleration is
certain to have a great
impact on our understanding of the ``dark energy'' and of the
underlying fundamental questions.

Type Ia supernovae can be used as standard candles to infer the luminosity
distance ($d_L$) as a 
function of redshift
\cite{Perlmutter:97,Riess:98,Perlmutter:99a}, and such data provide a key 
element in the case for cosmic acceleration.
By analyzing a simulated dataset, as might be obtained by
the proposed SNAP satellite\cite{SNAP}, we can test the 
ability of such experiments to distinguish among currently attractive models.
Our work will be reported more extensively in
\cite{bigone}. We note that where our calculations overlap we are in
complete numerical agreement with the recent paper by Maor {\it et al.}
\cite{Maor:00}.  However our perspective and emphasis is considerably
different. 

We consider a future supernova dataset that
has been converted to a table of effective magnitudes and
redshifts of objects with a single fiducial absolute magnitude $M$.  We will
consider both statistical and systematic uncertainties in the
magnitudes.  Typically the redshifts are known to sufficiently high
precision that their uncertainties can be ignored.

The magnitudes and redshifts are related by the luminosity distance 
according to 
\begin{equation}
	m(z) = {\cal M} + 5\log {\cal D}_L,
\label{mz}
\end{equation}
where 
\begin{equation}
{\cal D}_L = H_0d_L = H_0(1 + z)\int_0^z {c \over H\left(z^\prime\right)} 
\,dz^\prime
\label{distL}
\end{equation}
and ${\cal M} \equiv M  - 5 \log{H_0} + 25$.  When nearby supernovae
are used to determine ${\cal M}$ the uncertainties in the Hubble constant
$H_0$ drop out of Eqn. \ref{mz}.  It is through ${\cal D}_L(z)$ that specific
models of the evolution of the universe enter the picture. 
If we had just a few well-specified models of dark energy to compare,
then their predictions for ${\cal D}_L$ could be fit to the 
supernova dataset and their relative likelihoods determined.
Even if a model had a free parameter, that parameter could be  determined
to some accuracy. 
However, it is not understood what fundamental physical principles 
specify the actual dark energy in the Universe, so we are left with a
potentially infinite set of families of theoretical models to compare,
each
with a potentially large number of free parameters.
No finite dataset will ever select among all of these theoretical models
(see, for example, Maor {\it et al.})\cite{Maor:00}, who show nine models, from this
infinite set, that cannot be distinguished by a proposed supernova dataset).

In the face of this theoretical uncertainty it is useful to select a fitting
function whose parameters can characterize the $m(z)$ dataset independent of
any specific theory.  By choosing a function that also fits most extant
theories' ${\cal D}_L$ predictions well, with the fewest possible parameters,
we can then use these parameters as indices that label 
the theories so they can be compared with the dataset.
Although it is also desirable for the fitting parameters to represent physically
meaningful concepts of the current meta-theories, this is not necessary.
This approach has already been undertaken
by several authors in the context of supernova
datasets\cite{Perlmutter:99b,Huterer:99,Saini:99,Maor:00}. Each of these papers
uses different types of functions to fit ${\cal D}_L(z)$.  Here we describe a
reason to choose one of these functions.

Original reports of  supernova data used (two-parameter) models
based on adding a cosmological constant term to Einstein's
equations. This approach 
was generalized in \cite{Perlmutter:99b} to an effective, 
but constant, equation of state factor. 
Huterer and Turner \cite{Huterer:99} modeled the luminosity distance 
according to the simple power law $d_L(z)=\sum_{i=1}^{N}c_i z^i$.
Saini {\it et al.} \cite{Saini:99} used the form\footnote{
As we  completed this work a preprint came out using a fitting
function with even more parameters\cite{Chiba:00}.  We will discuss this function in \cite{bigone}
}
\begin{equation}
{{\cal D}_L(z)} = {{2 (1+z)}} \left[ { z - \alpha\sqrt{1+z} + \alpha
\over \beta z + \gamma \sqrt{1+z} + 2 - \alpha - \gamma }
\right]
.
\label{Sainieqn}
\end{equation}

Our preferred fitting function is motivated by cosmic 
acceleration driven by an extra 
energy component (``quintessence'' or ``dark energy'') with the
pressure $p$ and density $\rho$ related by an equation of state $p =
w\rho$. The cosmological constant case is reproduced
when $w = -1$. For more general constant values of $w$ this
equation of state implies $\rho \propto (1 + z)^{3{(w + 1)}}$  
To allow for a $z$-dependence of $w$, it has been expanded
 in a
power series in $z$ (as in \cite{Huterer:00,Maor:00}) or,equivalently,
in $(1+z)$\cite{bigone}: 
\begin{equation}
w(z) =  \sum_{i=0}^{N}w_i z^i =
   \sum_{i=0}^{N}\tilde{w}_i\left(1+z\right)^i.
\end{equation}
We assume a flat universe (theoretically preferred by inflation
and currently favored by CMB data). Thus the remaining parameter
required to fix the cosmology is $\Omega_m$, the density of ordinary
(pressureless) matter today in units of the critical density.
For these models we get
\begin{eqnarray}
{\cal D}_L(z)& = 
& {c(1+z)} \int\limits_0^z 
\left(1+z^\prime\right)^{-3/2}
\Bigg[
\Omega_{\rm m} +\Omega_Q \left(1+z^\prime\right)^{3\tilde{w}_0} \nonumber \\
& & \left.\times\exp\left\{3\sum\limits_{i=1}^N\frac{\tilde{w}_i}{i}
\left[
\left(1+z^\prime\right)^i-1
\right]
\right\}
\right]^{-1/2}\, 
dz^\prime \,  
\end{eqnarray}
where for a flat universe $\Omega_Q = 1 - \Omega_{\rm m} $. 
Equation ~5 takes on a simple form in terms of
$\tilde{w_i}$, but throughout the rest of this {\em letter} we use the
more intuitive quantity $w_i$.

\begin{figure}[h]
\centerline{\psfig{file=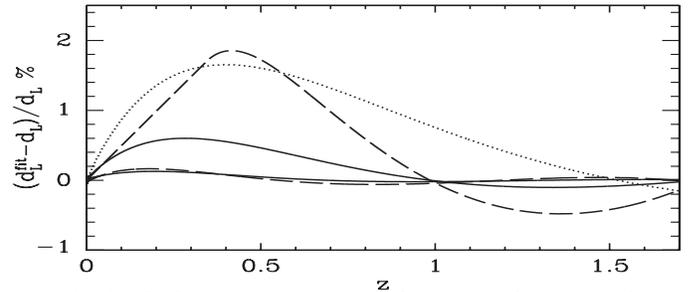,width=3.5in,height=1.5in}}
\caption{Residuals from three fitting expansions to a given model (the periodic potential given in
\protect\cite{Dodelson:00b}). The dashed lines are the quadratic
and cubic polynomial $d_L$-expansion fit, the solid lines are the linear and quadratic
$w$-expansion fit and the dotted line the fit according to
Eqn. \ref{Sainieqn}.
}
\label{dlfit}
\end{figure}

In
\cite{bigone} we survey essentially all the published quintessence
models, and in every case the $w$-expansion gave a better fit than
either Eqn. \ref{Sainieqn} or the power law expansion of $d_L$, when
taken to the same order of fit variables (see Fig.~\ref{dlfit}).
The quadratic order $d_L$-expansion (top dashed line) results in a
worse fit than the linear order $w$-expansion (top solid line), and
the quadratic order $w$-expansion (lower solid line) fits as well as 
the cubic order $d_L$-expansion (lower dashed line).
(This
result also holds when we weight the data points according to the
dataset specification described below, although all the fits improve.)

The $w$-expansion 
parameters also have an intuitive meaning in their own right.  Given
the uncertainty regarding causes of cosmic
acceleration, it might be reasonable to simply think of the $w_i$'s as 
the fundamental parameters for now.  They tell us something
relatively straightforward about the nature of the dark energy.
The same cannot be said of the $c_i$ of \cite{Huterer:99} or the
$\alpha, \beta$ and 
$\gamma$ of \cite{Saini:99}.  We are fortunate that the twin criteria 
of intuitive interpretation of parameters and good fits to published
models point to the same choice of fitting functions.  It is worth
noting however that we have not proven that the $w$-expansion provides 
the {\em best } possible fits to current models.  It is possible that
an even better choice could be discovered in the future.  

Maor {\it et al.} \cite{Maor:00} also favor the same $w$-expansion
formalism.  However we do not agree completely with their discussion
of the alternative methods \cite{Huterer:99,Saini:99}. Despite
claims to the contrary in \cite{Maor:00} all the
approaches use ``fitting functions''.  If for example a new
understanding of dark energy emerged in which the $c_i$ of \cite{Huterer:99} were
fundamental parameters, then the $w$-expansion would provide a bad fit
to those models and would be a bad choice for the fitting
functions (whereas $d_L(z)=\sum_{i=1}^{N}c_i z^i$ would be an ``exact
expression''). 

So we have argued that the $w$-expansion offers a good means of
interpreting supernova data.  Now we apply this expansion to simulated 
future supernova datasets.

\begin{figure}[!h]
\centerline{\psfig{file=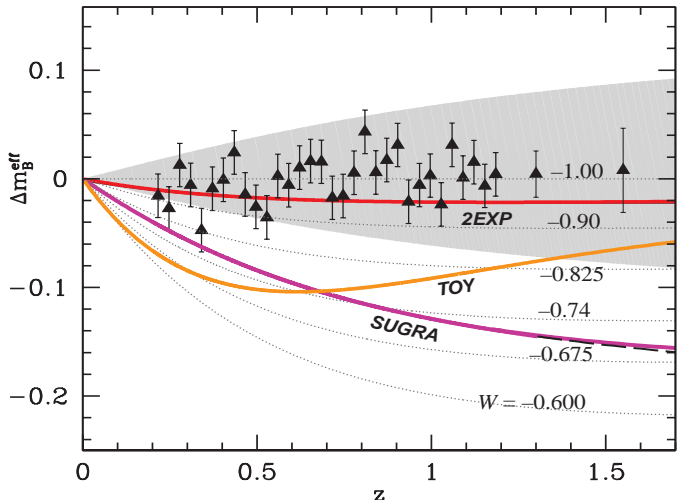,height=2.6in,width=3.5in}}
\caption{Simulated binned data, 
with solid curves from a supergravity
inspired model \protect\cite{Brax:99} (with the 
best $N=1$ fit shown as a dashed line, mostly covered by the 
solid {\it SUGRA} line), 
a strongly evolving toy model with
$w_0=-0.6$ and $w_1=-0.8$ ({\it TOY}), and a
model involving two exponentials 
\protect\cite{Barreiro:00} ({\it 2EXP}). A range of constant
$w$ models is shown as a grid of lines. The shaded region corresponds to cosmological constant
models with $0.25 \le \Omega_{\rm m} \le
0.35$ and $\Omega_\Lambda = 1-\Omega_{\rm m}$.}
\label{dmw0}
\end{figure}

For concreteness, we assume
a future dataset similar to one proposed for the SNAP satellite 
project\cite{SNAP},
with SNe observed uniformly within four different redshift ranges 
with the following different sampling rates: 
In the first range from $z=0-0.2$ we
assume that their are $50$ observations, in the second and largest
redshift range from $z=0.2-1.2$ there are $1800$ SNe and in the two
high redshift bins, $z=1.2-1.4$ and $z=1.4-1.7$, there are $50$ SNe and
$15$ SNe observations respectively. The statistical error in magnitude 
is assumed to be $\sigma_{\rm mag} = 0.15$, including both measurement
error and any residual
intrinsic dispersion after calibration.  It is worth noting that
considerations such as those raised in \cite{Huterer:00} could change
this redshift-distribution 
strategy to further optimize the impact of the data.
We construct Monte Carlo simulated datasets, 
first for a flat, cosmological
constant model with $\Omega_{\rm m}=0.3$ and $\Omega_\Lambda = 0.7$.  
Figure \ref{dmw0} shows an example of simulated data.  The points
show binned data with statistical error bars, where we combine
$56$ SNe in each bin.

Figure \ref{dmw0} shows the magnitude difference, $\Delta m$,
between several dark energy models (solid) and the Monte 
Carlo fiducial model, along with the 
simulated binned data for that model.
To guide the eye we also show a grid of constant $w$
models (dotted). 
Note how for the supergravity model
\cite{Brax:99} (labeled ``SUGRA''),
adding the linear term, $w_1$, improves the fit
considerably (as shown by the dashed line, mostly covered by the solid line), 
compared to the best constant fit with $w_0=0.74$. For
the two exponential model \cite{Barreiro:00} (``2EXP'') the constant fit seems
already sufficient. The line labeled ``TOY'' corresponds to a 
toy model with $w_0= -0.6$ and $w_1 = -0.8$, 
which is not well represented
by a constant fit.  All the models have $\Omega_m=0.3$.

Figure \ref{dmw0} illustrates that these models can be easily
 
differentiated from one another by the simulated data, and that for
the toy model the contribution
 
from $w_1$ looks large enough to be discriminated by the
data.  The targeted SNAP systematic uncertainties can
be estimated by considering for
example a linear drift up or down in the data (as a function of $z$)
which spans one error bar height over $0 < z < 1.5$  (Note that a 
constant systematic uncertainty would not affect the results.)
Further investigation is required to determine if these targets are
realistic. When the targeted
systematic uncertainties are taken into account the ``2EXP'' and
$w=-1$ models can no longer be discriminated.
In \cite{bigone} we report more thoroughly on the models which can and
cannot have $w_1$ differentiated from zero by such datasets. In most
published models $w_1$
  is below the level observable by supernovae
alone. 

We shall see below that the effect of
introducing $w_1$ in the fit for this dataset is to substantially {
increase} the uncertainty in $w_0$ (as emphasized in \cite{Maor:00}).
This is because with the additional fitting parameter there are many
more ways of fitting the simulated data to good accuracy.
This fact has {\em nothing} to do
with the ability of the (in this case simulated) dataset to
discriminate among the models shown.
The ability to discriminate among the models is a fundamental
property of the  dataset, whereas the choice of fitting function is a
matter of choosing suitable analysis techniques.  
As with any experimental data,
for a given
dataset each of the shown models
is a member of a degenerate family of 
models that cannot be differentiated 
by the data. Opening up the space of fitting functions allows
one to probe this degeneracy.  Furthermore, we show below how
realistic improvements in our independent
knowledge of $\Omega_m$ can break this degeneracy to an
interesting degree.

\begin{figure}[!h]
\centerline{\psfig{file=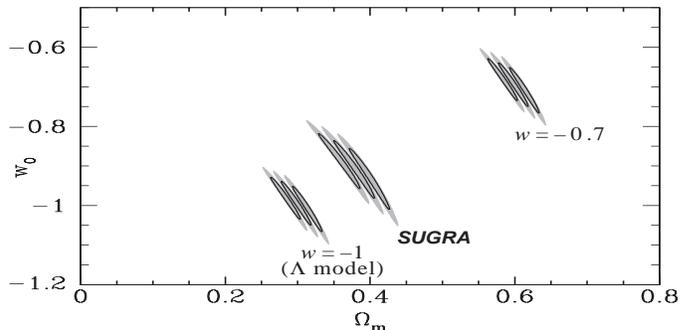,height=1.7in,width=3.5in}}
\caption{Separation of dark energy models (labeled as in Fig. 2)
in the $\Omega_{\rm m}$--$w_0$ plane.  We show
$39.3\%$ ($\Delta\chi^2 = 1$, shown as solid-line contours) 
and 68\% ($\Delta\chi^2 = 2.3$, shown as shaded contours) 
confidence regions which project on the axes to give
marginalized $1 \sigma$ and $1.5 \sigma$ errors respectively.  
The left and right flanking contours for each model show the
effects of a drifting systematic error (see text).  The $w=-0.7$
model has $\Omega_m=0.6$, the other two models are exactly the same as their
counterparts in Fig. 2.
More examples are given in \protect\cite{bigone}.}
\label{w0om}
\end{figure}

Figure \ref{w0om} shows joint likelihood contours in the $\Omega_m -
w_0$ plane for the simulated data, when these are the only parameters 
used to fit the data. We show also the shift due to a linearly 
drifting systematic
error of $\pm 0.02 \,{\rm mag}$ per 1.5 units in redshift.  Clearly the data 
will discriminate among the three models shown.

\begin{table}[!h]
\centering
\begin{tabular}{l c | cc}
prior $\sigma_{\Omega_m}$ & measurement $\sigma_{\rm mag} $ & $\sigma_{w_0}$ 
& $\sigma_{w_1}$ \\
\hline
No ${\Omega_m}$ prior; $w_1 = 0$& 0.15 & 0.06 &   \\
0.15 & 0.15 & 0.15 & 0.6 \\
0.05 & 0.15 & 0.06 & 0.2 \\
\,\,\," & {\it 0.09} & {\it 0.05} & {\it 0.12} \\
0.04 & 0.15 & 0.05 & 0.16 \\
0 \,\,\, (fixed ${\Omega_m}$) & 0.15 & 0.03 & 0.12 \\
\end{tabular}
\caption{Statistical measurement 
uncertainties on $w_0$ (i.e., $w_{\rm today}$) and $w_1$, given 
supernova magnitude measurement uncertainty, $\sigma_{\rm mag}$, and 
a range of uncertainties, $\sigma_{\Omega_m}$, in the 
independent prior knowledge of $\Omega_m$.   (As in Fig. 4, the supergravity
model is used here as the example, but the other models give comparable 
results.)  The systematic uncertainties
are 0.01 in $w_0$ and 0.10 in $w_1$.
}
\end{table}

\begin{figure}[!h]
\centerline{\psfig{file=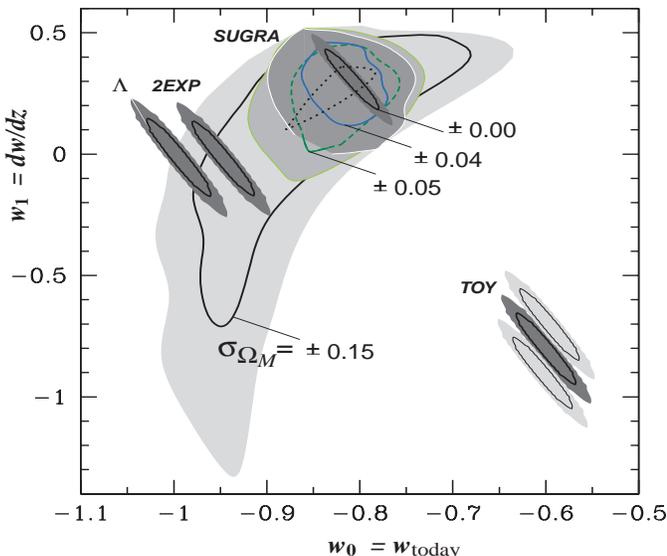,height=2.9in,width=3.5in}}
\caption{
Separation of dark energy models (labeled as in Fig. 2)
in the $w_0$--$w_1$ plane.  The shaded regions are $39.3\%$ joint
probabilities for the different $\Omega_{\rm m}$ priors and the dashed
and solid lines the corresponding $68\%$ probabilities.
For the example of the supergravity ({\it SUGRA}) model, a range of increasingly
larger contours represents the result using prior knowledge on $\Omega_m$ with
increasingly poorer uncertainty, $\sigma_{\Omega_m}$.
The black
dotted curve corresponds to the projected-1$\sigma$ error contour 
obtained for more optimistic dataset 
specifications ($\sigma_{\rm mag} = 0.09$; see text) with a prior of
$0.25\le\Omega_m\le0.35$.  
For the example of the {\it TOY} model, flanking contours show the
effects of a drifting systematic error (see text).}  
\label{w1w0}
\end{figure}

We now analyze the same simulated data using a fitting function that
includes $w_1$.  For this analysis (results shown in Table 1 and
Figure \ref{w1w0}), we consider
different possible states of prior information about the value of $\Omega_m$,
ranging from relatively poorly known ($\sigma_{\Omega_m} \approx 0.15$), roughly
the case today, to well determined ($\sigma_{\Omega_m} \approx 0.04$), 
a possible
goal within the next decade.  
We show in \cite{bigone} that the systematic errors
lead to roughly the same
contribution as the statistical errors for the case of a perfectly constrained  
$\sigma_{\Omega_m} = 0.0$.
We conclude from Fig. \ref{w1w0} and Table 1 that expanding the fitting function
to include $w_1$ will become useful when the value of
$\Omega_m$ is better constrained than it is today.  This result
adds to the case for producing an independent
determination of $\Omega_m$.

It may be possible to reduce the statistical uncertainties
even further, by increasing the sample size and/or by improving
the intrinsic luminosity calibration for each supernova.
Figure \ref{w1w0} also
shows the smaller likelihood region (dotted ellipse) which one gets
when taking $\sigma_{\rm mag} = 0.09$ and doubling the number of supernovae in 
each bin. 
This means (see Table 1) that the $1\sigma$ statistical error
on $w_1$ improves from $\sigma_{w_1} = 0.6$ for the current constraints
on $\Omega_m$ (solid line in Fig.4), to $\sigma_{w_1} = 0.2$ for future 
$\Omega_m$ constraints (dashed line), to very tight
constraints of $\sigma_{w_1} = 0.12$ for the ``improved'' 
scenario, with twice as many more-tightly-calibrated supernovae and
$\sigma_{\rm mag} = 0.09$ (dotted line). 

Figure \ref{w1w0} is similar to Fig. 2 from
\cite{Maor:00}. (To compare actual numbers for predicted uncertainties, note
that Maor {\it et al.} quote the full-width range of 95\%-confidence 
contours, i.e. four times a 1$\sigma$ uncertainty. Also, our different
contour shapes are due to our choice of a Gaussian probability
distribution for $\Omega_m$ vs. their tophat distribution.)   Maor {\it
et al.} emphasize the uncertainties that emerge when the fitting
function is expanded to include $w_0$ and $w_1$ given our current
knowledge of $\Omega_m$. We disagree with this emphasis and focus
instead on the great impact 
a SNAP-type experiment will have.
Even if our determination of $\Omega_m$ 
were not to improve over the next decade, 
SNAP-class datasets would still be powerful
discriminators among models. For example, such a dataset could easily
discriminate between $w=-1$ (a cosmological constant) as the source of cosmic
acceleration and several interesting quintessence models (such as
SUGRA or others discussed in \cite{bigone}). A result either way would
have profound
 
implications.
Furthermore, significant improvements on the determination
of $\Omega_m$ are expected over the next decade and we have
shown how sufficient improvements would allow the extra fitting
parameter $w_1$ to become an ``observable'' and help to further
differentiate theories \cite{concerns}.

We have investigated different analysis methods for supernova datasets,
and used our preferred method to explore the prospects of further
supernova searches.  It 
is clear that a dataset of the sort proposed with 
SNAP \cite{SNAP} presents an exciting opportunity to 
constrain theories of cosmic acceleration.

ACKNOWLEDGMENTS: 
We thank S. Perlmutter for extensive discussions and in particular for
emphasizing the importance of the improving knowledge of $\Omega_m$; 
D. Huterer, A. Lewin, P. Steinhardt and M. Turner for helpful
conversations; and
 
P. Astier for comparing our results with his independent calculations.
We were supported by DOE grant DE-FG03-91ER40674
and U.C. Davis.

\vspace{-.2in}

\def\jnl#1#2#3#4#5#6{\hang{#1, {\it #4\/} {\bf #5}, #6 (#2).}}
\def\jnltwo#1#2#3#4#5#6#7#8{\hang{#1, {\it #4\/} {\bf #5}, #6; {\it
ibid} {\bf #7} #8 (#2).}} 
\def\prep#1#2#3#4{\hang{#1, #4.}} 
\def\proc#1#2#3#4#5#6{{#1, in {\it #3 (#4)\/}, edited by #5,\ (#6).}}
\def\book#1#2#3#4{\hang{#1, {\it #3\/} (#4, #2).}}
\def\jnlerr#1#2#3#4#5#6#7#8{\hang{#1 [#2], {\it #4\/} {\bf #5}, #6.
{Erratum:} {\it #4\/} {\bf #7}, #8.}}
\def\prl{Phys.\ Rev.\ Lett.}
\def\pr{Phys.\ Rev.}
\def\pl{Phys.\ Lett.}
\def\np{Nucl.\ Phys.}
\def\prp{Phys.\ Rep.}
\def\rmp{Rev.\ Mod.\ Phys.}
\def\cmp{Comm.\ Math.\ Phys.}
\def\mpl{Mod.\ Phys.\ Lett.}
\def\apj{Ap.\ J.}
\def\apjl{Ap.\ J.\ Lett.}
\def\aap{Astron.\ Ap.}
\def\cqg{Class.\ Quant.\ Grav.} 
\def\grg{Gen.\ Rel.\ Grav.}
\def\mn{MNRAS}
\def\ptp{Prog.\ Theor.\ Phys.}
\def\jetp{Sov.\ Phys.\ JETP}
\def\jetpl{JETP Lett.}
\def\jmp{J.\ Math.\ Phys.}
\def\zpc{Z.\ Phys.\ C}
\def\cupress{Cambridge University Press}
\def\pup{Princeton University Press}
\def\wss{World Scientific, Singapore}
\def\oup{Oxford University Press}
\def\asj{Astron.~J}

\pagebreak
\pagestyle{empty}

\end{document}